\documentclass[onecolumn,showpacs,preprintnumbers,amsmath,amssymb]{revtex4}
\usepackage[a4paper,left=2cm,right=2cm,top=2cm,bottom=3cm]{geometry}
\usepackage{amsmath}
\usepackage{amssymb}
\usepackage{graphicx}
\usepackage{bm}
\usepackage{epstopdf}
\usepackage{slashed}

\begin{document}

\title {Material boundaries in Carroll-Field-Jackiw Lorentz-violating electrodynamics}

\author{David M. Soares}
\email{david.junior@estudante.ufla.br}

\author{L. H. C. Borges}
\email{luizhenrique.borges@ufla.br}

\author{G. Dallabona}
\email{gilson.dallabona@ufla.br}

\author{L. C. T. Brito}
\email{lcbrito@ufla.br}

\affiliation{Departamento de Física, Universidade Federal de Lavras,
Caixa Postal 3037, 37200-900 Lavras-MG, Brazil}

%\author{F.A. Barone}
%\email{fbarone@unifei.edu.br}
%\affiliation{IFQ - Universidade Federal de Itajub\'a, Av. BPS 1303, Pinheirinho, Caixa Postal 50, 37500-903, Itajub\'a, MG, Brazil}

%\author{F.E. Barone}
%\email{febarone@cbpf.br}
%\affiliation{Niter\'oi, RJ, Brazil}

%%%%%%%%%
\begin{abstract}

This paper investigates certain aspects of the CPT-odd photon sector of the minimal Standard Model Extension (SME) in the presence of a perfectly conducting plate (perfect mirror). The considered sector is described by Carroll-Field-Jackiw (CFJ) electrodynamics, where Lorentz violation is due to the presence of a single background vector denoted as $\left(k_{AF}\right)^{\mu}$, which we treat perturbatively up to second order. Specifically, we derive the modified propagator for the electromagnetic field due to the presence of the perfect mirror, and we study the corresponding interaction between the mirror and a stationary point-like charge. Our results reveal that when the charge is positioned near the mirror, a spontaneous torque emerges, which is a unique effect of Lorentz symmetry breaking. Additionally, we demonstrate that the image method for this theory is applicable when the background vector has only the component perpendicular to the mirror.

\end{abstract}
%%%%%%%%%

\maketitle

%%%%%%%%%
\section{Introduction}
\label{I}
%%%%%%%%%

Lorentz violation theories have been intensively investigated as a possibility to provide information on new physics at the Planck scale. Such investigations have mainly been carried out within the minimal Standard Model Extension (SME) \citep{SME1,SME2,SME3}. In particular, the electromagnetic sector of the minimal SME is composed by a CPT-odd and a CPT-even part. The CPT-odd part is described by the Carroll-Field-Jackiw (CFJ) electrodynamics, where the Maxwell
Lagrangian is augmented by the term $\sim\epsilon^{\mu\nu\alpha\beta}\left(k_{AF}\right)_{\mu}A_{\nu}F_{\alpha\beta}$ \citep{CFJ1}, with $\left(k_{AF}\right)^{\mu}$ being the background vector responsible for the Lorentz violation.  The effects of the CFJ term have been extensively studied in the literature with respect to classical electrodynamics \citep{CL1,CL2,CL3,CL4,CL5}, radiative corrections \citep{R1,R2,R3,R4,R5}, quantum electrodynamics (QED) \citep{QED1,QED2}, topological defects  \citep{TP1,TP2}, and many others. The CPT-even sector is obtained by adding the term $\sim\left(K_{F}\right)^{\mu\nu\alpha\beta}F_{\mu\nu}F_{\alpha\beta}$ in the usual Maxwell Lagrangian, where the background tensor $\left(K_{F}\right)^{\mu\nu\alpha\beta}$ exhibits 19 independents coefficients, with 10 of them being sensitive to birefringence and 9 being nonbirefringent \citep{Ce0}. Some aspects involving this sector were investigated in \citep{Ce1,Ce2,Ce3,Ce4,Ce5,Ce6,Ce7,Ce8,Ce9,Ce10,Ce11,Ce12,Ce13}.

Quantum field theories in the presence of nontrivial boundary conditions have been a subject of great interest and exploration in the literature, finding vast applications in several branches of physics. Examples include studies where $\delta$-like potentials coupled to fields are used to describe material boundaries \citep{FABFEB, GTFABFEB, OliveiraBorgesAFF, Milton, BorUM, KimballA, BordKD, NRVMH, NRVMMH2, PsRj, Caval, FABFEB2}, investigations related to Lee-Wick electrodynamics \citep{FABAAN1, LW1, LW2, LW3, BorgesBarone22}, and the examination of certain aspects of planar theories due to the presence of boundary conditions \cite{BorgesBarone22, plane1, plane2}.

Theories involving Lorentz symmetry breaking under the influence of boundary conditions warrant further investigation, as electromagnetic configurations in actual experiments are typically surrounded by conductors. These conductors need to be properly considered in theoretical models. In this context, several investigations have been carried out, including those concerning Casimir energy \citep{CSE1, CSE2, CSE3, CSE4, CSE5, CSE6, CSE7, CSE8, CSE9, CSE10, CSE11, CSE12, CSE13, CSE14, CSE15, CSE16, CSE17, CSE18, CSE19, CSE20}. Additionally, studies have explored Lorentz-violating Maxwell electrodynamics due to the presence of a perfect conductor (perfect mirror) in the minimal SME \citep{LHCBFABplate} and in the nonminimal SME \citep{LHCBFABplate2}, which includes Lorentz-violating terms containing higher derivatives \cite{H1, H2, H3}, as well as effects related to the presence of a semi-transparent mirror \citep{LHCBAFFFAB, LHCBAFFSM}. Focusing on the minimal SME, Ref. \cite{LHCBFABplate} investigated some features of the Lorentz-violating CPT-even gauge sector in the presence of a perfectly conducting plate. However, a similar investigation for the CPT-odd gauge sector has not been considered in the literature until now. Exploring this aspect could be of interest to understanding the types of physical phenomena that may arise in such a sector in the vicinity of a conductor. Moreover, there has been significant interest in understanding the features of CFJ electrodynamics.

This paper is dedicated to exploring   the effects of a perfectly conducting plate on a charged particle in the context of the CPT-odd pure-photon sector of the minimal SME. Specifically, we focus on the CFJ model, investigating Lorentz violation effects arising from the presence of a single perfect mirror. The background vector $\left(k_{AF}\right)^{\mu}$ is assumed to be very small, and we treat it perturbatively up to second order. In Section \ref{II}, we derive the propagator for the gauge field in the presence of the conducting plate. In Section \ref{III}, we use this propagator to calculate the interaction energy and interaction force between a stationary point-like charge and the perfect mirror. We show that placing the charge near the mirror induces a spontaneous torque in the system, a phenomenon absent in usual Maxwell electrodynamics with a perfect mirror. We also compare the results with the free theory (theory without the plate) and verify that the image method is valid when the background vector has only the component perpendicular to the mirror. Section \ref{IV} is dedicated to our final remarks and conclusions.

Throughout the paper we work in a $(3+1)$   Minkowski space-time with metric $\eta^{\rho\nu}=(1,-1,-1,-1)$.

%%%%%%%%%
\section{THE PROPAGATOR IN THE PRESENCE OF A MATERIAL BOUNDARY}
\label{II}
%%%%%%%%%

The CFJ model is part of the electromagnetic CPT-odd sector of the minimal SME and is described by the following Lagrangian density \citep{SME3,CFJ1}, 
%%%%%%%%%%%%%%%%%%%%%%%%%
\begin{equation}
{\cal L}=-\frac{1}{4}F_{\mu\nu}F^{\mu\nu}-\frac{1}{2\gamma}\left(\partial_{\mu}A^{\mu}\right)^{2}+\frac{1}{2}\epsilon^{\mu\nu\alpha\beta}\left(k_{AF}\right)_{\mu}A_{\nu}F_{\alpha\beta}-J^{\mu}A_{\mu}\ .\label{lagEm}
\end{equation}
%%%%%%%%%%%%%%%%%%%%%%%
Here $A^{\mu}$ is the photon field, $F^{\mu\nu}=\partial^{\mu}A^{\nu}-\partial^{\nu}A^{\mu}$
stands for the field strength, $J^{\mu}$ is an electromagnetic field source and $\gamma$ is a gauge fixing parameter. The constant background vector $\left(k_{AF}\right)^{\mu}$ stands for the Lorentz violation parameter, with mass dimension one. Given that the background vector is significantly smaller than any relevant physical scale in the problem, throughout the paper, we treat it perturbatively up to second order.

The model (\ref{lagEm}) can be rewrite in the following way
%%%%%%%%%%%%%%%%
\begin{equation}
\label{lagempe}   
{\cal L}\rightarrow\frac{1}{2}A_{\mu}{\cal{O}}^{\mu\nu}A_{\nu}-J^{\mu}A_{\mu} \ ,
\end{equation}
%%%%%%%%%%%%%%%%%%%%
where we defined the differential operator 
%%%%%%%%%%%%%%%%% 
 \begin{equation}
{\cal{O}}_{\mu\nu} =\Box\eta_{\mu\nu}-\left(1-\frac{1}{\gamma}\right)\partial_{\mu}\partial_{\nu}-2\epsilon_{\mu\nu\rho\alpha}\left(k_{AF}\right)^{\rho}\partial^{\alpha}  \ . 
 \end{equation}
%%%%%%%%%%%%%%%%%%%%%%%
 
 The propagator $D^{\mu\nu}\left(x,y\right)$ satisfies, 
%%%%%%%%%%%%%%%%%%
\begin{equation}
\label{prop1}
{\cal{O}}_{\mu\nu}D^{\nu}_{\ \lambda}\left(x,y\right)=\eta_{\mu\lambda}\delta^{4}\left(x-y\right) \ .
\end{equation}
 %%%%%%%%%%%%%%%
 
 By using the Feynman gauge $\gamma=1$, one can solve the above equation obtaining the propagator up to second order in the Lorentz-violating parameter $\left(k_{AF}\right)^{\mu}$ \cite{CL1}, 
%%%%%%%%%%%%%%%%%
\begin{eqnarray}
 \label{prop2} 
D^{\mu\nu}\left(x,y\right)&=&-\int\frac{d^{4}p}{\left(2\pi\right)^{4}}\frac{e^{-ip\cdot\left(x-y\right)}}{p^{2}}\Biggl\{\left[1+\frac{4\left[\left(k_{AF}\right)\cdot p\right]^{2}}{p^{4}}-\frac{4\left(k_{AF}^{2}\right)}{p^{2}}\right]\eta^{\mu\nu}+\frac{4\left(k_{AF}^{2}\right)}{p^{2}}\frac{p^{\mu}p^{\nu}}{p^{2}}\nonumber\\
&
&+\frac{2i}{p^{2}}\epsilon^{\mu\nu\alpha\beta}\left(k_{AF}\right)_{\alpha}p_{\beta}
+\frac{4\left(k_{AF}\right)^{\mu}\left(k_{AF}\right)^{\nu}}{p^{2}}-\frac{4\left[\left(k_{AF}\right)\cdot p\right]}{p^{4}}\left[\left(k_{AF}\right)^{\mu}p^{\nu}+\left(k_{AF}\right)^{\nu}p^{\mu}\right]\Biggr\} \ . 
\end{eqnarray}
%%%%%%%%%%%%%%%%%%
   
Now, for the model (\ref{lagEm}), we consider the presence of a single perfectly conducting surface (perfect mirror). As discussed in Ref. \cite{LHCBFABplate}, the presence of such a surface in a Lorentz violation scenario imposes a boundary condition on the gauge field in such a way that the Lorentz force on it must be vanishing. This condition is given by
%%%%%%%%%%%%%%%%%%%%
\begin{equation}
\label{fdual}
n^{\mu}F^{*}_{\ \mu\nu}=0 \ ,
\end{equation}
%%%%%%%%%%%%%%%%
where $F^{*}_{\ \mu\nu}=(1/2)\epsilon_{\mu\nu\alpha\beta}F^{\alpha\beta}$ is the dual field strength, $\epsilon^{\mu\nu\alpha\beta}$ corresponds to the Levi-Civita tensor with $\epsilon^{0123}=1$ and $n^{\mu}$ is the four-vector perpendicular to the mirror. 

\begin{figure}[!h]
\centering 
\includegraphics[scale=0.40]{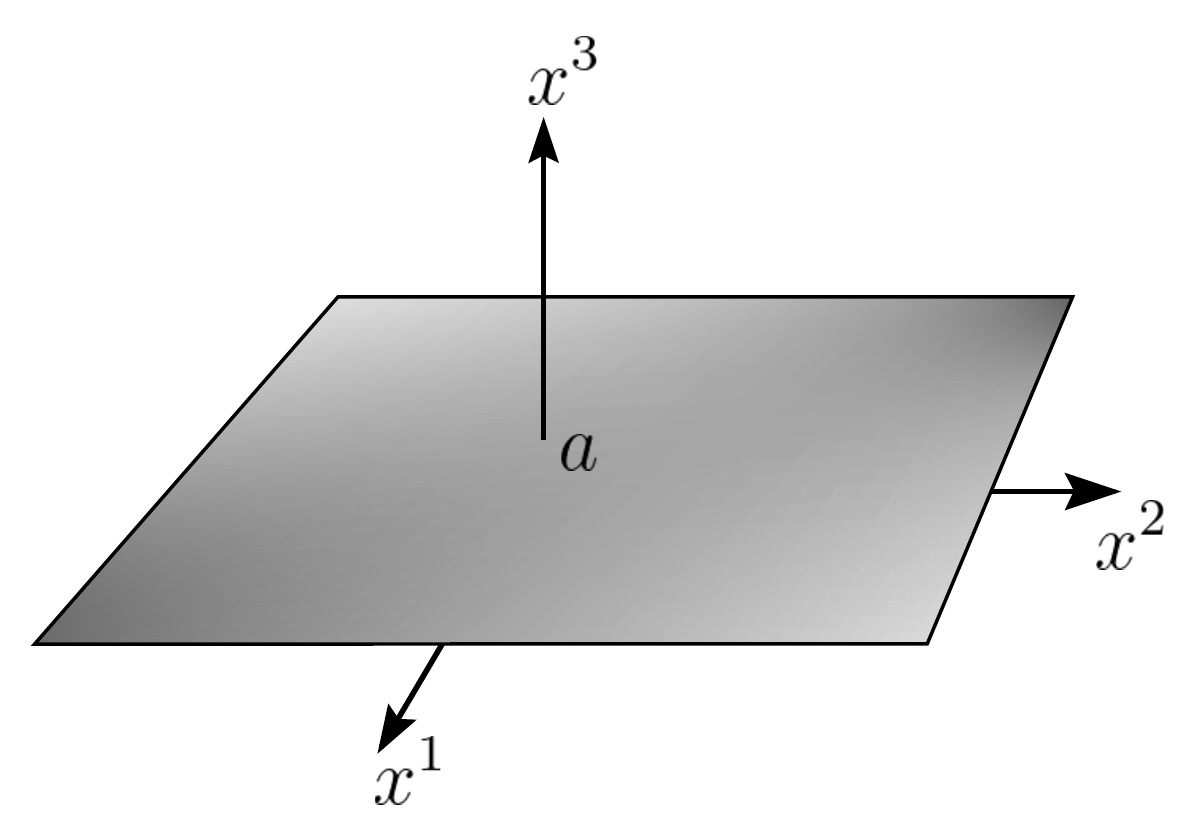}
\caption{Conducting surface placed on the plane $x^{3}=a$.}
\label{figu1}
\end{figure}

 Let us choose a coordinate system where the conducting plate is perpendicular to the $x^{3}$ axis, placed on the plane $x^{3}=a$ (see the Fig. \ref{figu1}), so that, $n^{\mu}=\eta^{\ \mu}_{3}=\left(0,0,0,1\right)$ is the four-vector normal to the surface. Therefore, the boundary condition (\ref{fdual}) reads
%%%%%%%%%%%%%%
\begin{eqnarray}
\label{condition1}
F^{*}_{\ 3\nu}\left(x\right)|_{x^{3}=a}=\epsilon_{3\nu}^{\ \ \alpha\beta}\partial_{\alpha}
A_{\beta}\left(x\right)|_{x^{3}=a}=0 \ ,
\end{eqnarray}   
%%%%%%%%%%%%%%%%
where the sub-index means that the boundary conditions are taken on the plane $x^{3}=a$.
%%%%%%%%%%%%%%%%%%%%%%%%%
%%%%%%%%%%%%%%%%%%%%%%%%%%%%%%%%%

In order to obtain the functional generator for the gauge field subordinated to the boundary conditions (\ref{condition1}) we will use the functional formalism employed in \citep{FABAAN1,BorgesBarone22,plane1,plane2,LHCBFABplate,LHCBFABplate2,MBor85}. The functional generator is given by
%%%%%%%%%%%%%%%    
\begin{eqnarray}
\label{fgen1}
Z_{C}\left[J\right]=\int {\cal{D}}A_{C} \ e^{i\int d^{4}x \ \cal{L}} \ ,
\end{eqnarray}
%%%%%%%%%%%%%%%%
where the sub-index $C$ means that we are integrating out in all the field configurations that satisfy the boundary conditions (\ref{condition1}). This restriction is reached by inserting, in the standard functional generator, a functional delta which is no-null only where the conditions (\ref{condition1}) are satisfied,  
%%%%%%%%%%%%%%%    
\begin{eqnarray}
\label{fgen2}
Z_{C}\left[J\right]=\int {\cal{D}}A \ \delta\left[F^{*}_{\ 3\nu}\left(x\right)
|_{x^{3}=a}\right] \ e^{i\int d^{4}x \ \cal{L}} \ .
\end{eqnarray}
%%%%%%%%%%%%%%%% 

Using the functional Fourier representation for the functional delta, one can write
 %%%%%%%%%%%%%%%    
\begin{eqnarray}
\label{fgen3}
\delta\left[F^{*}_{\ 3\nu}\left(x\right)|_{x^{3}=a}\right]=\int {\cal{D}}B\exp\left[i
\int d^{4}x\ \delta\left(x^{3}-a\right)B_{\nu}\left(x_{\parallel}\right)F^{*\ \nu}_{\ 3}
\left(x\right)\right] \ ,
\end{eqnarray}
%%%%%%%%%%%%%%%%  
with $B_{\nu}\left(x_{\parallel}\right)$ standing for an auxiliary vector field and
$x_{\parallel}^{\mu}=\left(x^{0},x^{1},x^{2}\right)$ means we have only the coordinates parallel to the surface.

We notice that the auxiliary field  $B_{\nu}\left(x_{\parallel}\right)$ exhibits gauge symmetry
 %%%%%%%%%%%%%%%    
\begin{eqnarray}
\label{gsy}
B^{\left(k\right)}_{\nu}\left(x_{\parallel}\right)\rightarrow B^{\left(k\right)}
_{\nu}\left(x_{\parallel}\right)+\partial_{\nu\parallel}\Lambda^{\left(k\right)}
\left(x_{\parallel}\right) \ .
\end{eqnarray}
%%%%%%%%%%%%%%%%
Thus, with the aim of discard the infinite gauge volume from the integral (\ref{fgen3}), we employ the
Faddeev-Popov method and the t'Hooft trick. We start by fixing the covariant gauge
%%%%%%%%%%%%%%%
\begin{equation}
\label{A21}
F\left[B_{\nu}\left(x_{\parallel}\right)\right]=\partial_{\parallel}^{\nu}
B_{\nu}\left(x_{\parallel}\right)=f\left(x_{\parallel}\right) \ ,
\end{equation}
%%%%%%%%%%%%%%%
where $f\left(x_{\parallel}\right)$ is an arbitrary function. Once the 
Faddeev-Popov determinant is independent of the field $B_{\nu}^{k}\left(x_{\parallel}\right)$,
it does not contribute to the integral. Therefore, we can write the functional delta (\ref{fgen3}) in the following way 
%%%%%%%%%%%%%%
\begin{eqnarray}
\label{A22}
\delta\left[F^{*}_{\ 3\nu}\left(x\right)|_{x^{3}=a}\right]\sim\int {\cal{D}}B
\ \delta\left[F[B_{\nu}(x_{\parallel})]-f\left(x_{\parallel}\right)\right]
\cr\cr
\times\exp\left[i\int d^{4}x\ \delta\left(x^{3}-a\right)B_{\nu}\left(x_{\parallel}\right)F^{*\ \nu}_{\ 3}
\left(x\right)\right] \ .
\end{eqnarray}
%%%%%%%%%%%%%%%

Now, we perform an integration by parts in the argument of the exponential and apply t'Hooft trick. We multiply both sides of Equation (\ref{A22}) by a convergent functional of $f\left(x_{\parallel}\right)$ and then integrate over $f\left(x_{\parallel}\right)$,
%%%%%%%%%%%%%%
\begin{eqnarray}
\label{A23}
\delta\left[F^{*}_{\ 3\nu}\left(x\right)|_{x^{3}=a}\right]=N\int {\cal{D}}f\int {\cal{D}}B
\delta\left[F[B_{\nu}(x_{\parallel})]-f\left(x_{\parallel}\right)\right]
\cr\cr
\times\exp\left[-i\int d^{4}x\ \delta\left(x^{3}-a\right)A_{\beta}\left(x\right)\epsilon_{3}^{\ \nu\alpha\beta}
\partial_{\alpha}B_{\nu}\left(x_{\parallel}\right)\right] 
\cr\cr
\times\exp\left[-\frac{i}{2\xi}\int d^{4}x \ d^{4}y
 \ \delta\left(x^{3}-a\right)f\left(x_{\parallel}\right)Q\left(x,y\right)f\left(y_{\parallel}\right)
\delta\left(y^{3}-a\right)\right] \ .
\end{eqnarray}
%%%%%%%%%%%%%%%
Here, $Q\left(x,y\right)$ represents an arbitrary function that must be chosen conveniently, $N$ is a constant, and $\xi$ is a gauge-fixing term.

Carrying out the functional integral in $f$ and some manipulations in Eq. (\ref{A23}), we have
%%%%%%%%%%%%%%
\begin{eqnarray}
\label{A24}
\delta\left[F^{*}_{\ 3\nu}\left(x\right)|_{x^{3}=a}\right]=N\int{\cal{D}}B \exp\left[-i\int d^{4}x\ 
\delta\left(x^{3}-a\right)A_{\beta}\left(x\right)\epsilon_{3}^{\ \nu\alpha\beta}\partial_{\alpha}B_{\nu}
\left(x_{\parallel}\right)\right]
\cr\cr
\times\exp\Bigl[-\frac{i}{2\xi}\int d^{4}x \ d^{4}y \ \delta\left(x^{3}-a\right)B_{\mu}
\left(x_{\parallel}\right)\frac{\partial^{2}Q\left(x,y\right)}{\partial x_{\mu\parallel}\partial 
y_{\nu\parallel}}\delta\left(y^{3}-a\right)B_{\nu}\left(x_{\parallel}\right)\Bigr] \ .
\end{eqnarray}
%%%%%%%%%%%%%%%

Substituting (\ref{A24}) in (\ref{fgen2}) we arrive at
%%%%%%%%%%%%%%
\begin{eqnarray}
\label{A25}
Z_{C}\left[J\right]=N\int {\cal{D}}A {\cal{D}}B \ e^{i\int d^{4}x \ \cal{L}}
\exp\left[-i\int d^{4}x\ 
\delta\left(x^{3}-a\right)A_{\beta}\left(x\right)\epsilon_{3}^{\ \nu\alpha\beta}\partial_{\alpha}B_{\nu}
\left(x_{\parallel}\right)\right]
\cr\cr
\times\exp\left[-\frac{i}{2\xi}\int d^{4}x \ d^{4}y \ \delta\left(x^{3}-a\right)B_{\mu}
\left(x_{\parallel}\right)\frac{\partial^{2}Q\left(x,y\right)}{\partial x_{\mu\parallel}\partial 
y_{\nu\parallel}}\delta\left(y^{3}-a\right)B_{\nu}\left(x_{\parallel}\right)\right] \ .
\end{eqnarray}
%%%%%%%%%%%%%%%

We notice that in the first exponential we have only the presence of $A^{\mu}$ field and only $B^{\mu}$ in the third one. However, the second exponential exhibits a coupling involving $A$ and $B$. To decouple these fields, we must perform the following translation
%%%%%%%%%%%%%%
\begin{eqnarray}
\label{A26}
A^{\beta}\left(x\right)\rightarrow A^{\beta}\left(x\right)+\int d^{4}y D^{\beta}_{\ \alpha}\left(x,y\right)
\delta\left(y^{3}-a\right)\epsilon_{3}^{\ \nu\gamma\alpha}\partial_{\gamma}B_{\nu}\left(y\right) \ ,
\end{eqnarray}
%%%%%%%%%%%%%%%
that enable us to write the expression (\ref{A25}) as follows
%%%%%%%%%%%%%%%%
\begin{eqnarray}
\label{fgen5}
Z_{C}\left[J\right]=NZ\left[J\right]{\bar{Z}}\left[J\right] \ ,
\end{eqnarray}
%%%%%%%%%%%%%%%%
where $N$ is a constant which is independent of fields, $Z\left[J\right]$ is the standard functional generator for the gauge field $A^{\mu}\left(x\right)$,
%%%%%%%%%%%%%%%    
\begin{eqnarray}
\label{fgen6}
Z\left[J\right]=\int{\cal{D}}A\ e^{i\int d^{4}x \ \cal{L}}
=Z\left[0\right]\exp\left[-\frac{i}{2}\int d^{4}x \ d^{4}y \ J^{\mu}\left(x\right)D_{\mu\nu}
\left(x,y\right)J^{\nu}\left(y\right)\right] \ ,
\end{eqnarray}
%%%%%%%%%%%%%%%%
and ${\bar{Z}}\left[J\right]$ stands for the contribution due to the vector field 
$B_{\nu}\left(x_{\parallel}\right)$,  
%%%%%%%%%%%%%%%    
\begin{eqnarray}
\label{fgen7}
{\bar{Z}}\left[J\right]=\int{\cal{D}}B\exp\left[i\int d^{4}x \ \delta
\left(x^{3}-a\right)I^{\nu}\left(x\right)B_{\nu}\left(x_{\parallel}\right)\right] \nonumber\\
\times\exp\left[-\frac{i}{2}\int d^{4}x \ d^{4}y \ \delta\left(x^{3}-a\right)
\delta\left(y^{3}-a\right)B_{\nu}\left(x_{\parallel}\right)W^{\nu\pi}\left(x,y\right)
B_{\pi}\left(y_{\parallel}\right)\right] \ ,
\end{eqnarray}
%%%%%%%%%%%%%%%%
where we identified
%%%%%%%%%%%%%%%    
\begin{eqnarray}
\label{defi1}
I^{\nu}\left(x\right)&=&-\int d^{4}y \ \epsilon_{3}^{\ \nu\gamma\alpha}
\left(\frac{\partial}{\partial x^{\gamma}}D_{\alpha\mu}\left(x,y\right)
\right)J^{\mu}\left(y\right) \ ,\\ 
\label{defi2}
W^{\nu\pi}\left(x,y\right)&=&\epsilon_{3}^{\ \nu\alpha\lambda}
\epsilon_{3}^{\ \pi\gamma\rho}\frac{\partial^{2}D_{\lambda\rho}\left(x,y\right)}
{\partial x^{\alpha}\partial y^{\gamma}}+\frac{1}{\xi}\frac{\partial^{2}Q\left(x,y\right)}
{\partial x_{\nu\parallel}\partial y_{\pi\parallel}} \ ,
\end{eqnarray}
%%%%%%%%%%%%%%%%

The integral (\ref{fgen7}) can be evaluated exactly. In order to accomplish this task, it is appropriate to perform the following choice
%%%%%%%%%%
\begin{eqnarray}
\label{QXY1}
Q(x,y)=-\int\frac{d^{4}p}{(2\pi)^{4}}e^{-ip\cdot(x-y)}\left[\frac{1}{p^{2}}+\frac{4\left[\left(k_{AF}\right)\cdot p\right]^{2}}{p^{6}}+\frac{4\left[\left(k_{AF}\right)^{3}\right]^{2}}{p^{4}}
\right] \ ,
\end{eqnarray}
%%%%%%%%%%%%   
and work in the gauge where $\xi=1$, being $\left(k_{AF}\right)^{3}$  the background vector component perpendicular to the plate. 
Substituting (\ref{defi1}) and (\ref{defi2}) into (\ref{fgen7}), using (\ref{prop2}), (\ref{QXY1}) and the fact that (see appendix)
%%%%%%%%%%%%%%%%%
\begin{eqnarray}
\label{intp31}
\int \frac{dp^{3}}{2\pi}\frac{e^{i p^{3}(x^{3}-y^{3})}}
{p^{2}}&=&-\frac{i}{2\Gamma} e^{i\Gamma\mid x^{3}-y^{3}\mid} \ , \\
\label{intp312}
\int \frac{dp^{3}}{2\pi}\frac{e^{i p^{3}(x^{3}-y^{3})}}
{p^{4}}&=&-\frac{1}{4 p_{\parallel}^{2}}\left(\frac{i}{\Gamma}+\mid x^{3}-y^{3}\mid\right) 
 e^{i\Gamma\mid x^{3}-y^{3}\mid} \ , \\
\label{intp313}
\int \frac{dp^{3}}{2\pi}\frac{e^{i p^{3}(x^{3}-y^{3})}}
{p^{6}}&=&\frac{1}{16 p_{\parallel}^{4}}\left[-3\mid x^{3}-y^{3}\mid+\frac{i}{\Gamma}\left(-3+p_{\parallel}^{2}\mid x^{3}-y^{3}\mid^{2}\right)\right]e^{i\Gamma\mid x^{3}-y^{3}\mid} \ ,
\end{eqnarray}
%%%%%%%%%%%%%%%%%
where $p^{3}$ means the momentum perpendicular to the plate, $\Gamma=\sqrt{p_{\parallel}^{2}}$, and
$p_{\parallel}^{\mu}=\left(p^{0},p^{1},p^{2}\right)$ is the momentum parallel to the plate, defining the parallel metric
%%%%%%%%%%%%%%%%%
\begin{eqnarray}
\label{etapar}
\eta_{\parallel}^{\mu\nu}=\eta^{\mu\nu}-\eta_{\ 3}^{\mu}\eta^{\nu 3} \ ,
\end{eqnarray} 
%%%%%%%%%%%%%%%%%
we arrive at
%%%%%%%%%%%%%%%%%%%%
\begin{eqnarray}
\label{fgplapar}   
{\bar{Z}}\left[J\right]=\int{\cal{D}}B\exp\left[i\int d^{3}x_{\parallel} \ I^{\nu}\left(x_{\parallel}\right)B_{\nu}\left(x_{\parallel}\right)\right] \nonumber\\
\times\exp\left[-\frac{i}{2}\int d^{3}x_{\parallel} \ d^{3}y_{\parallel} \ B_{\nu}\left(x_{\parallel}\right)W^{\nu\pi}\left(x_{\parallel},y_{\parallel}\right)
B_{\pi}\left(y_{\parallel}\right)\right] \ ,
\end{eqnarray}
%%%%%%%%%%%%%%%%%%%%%%%
where 
%%%%%%%%%%%%%%%%%%%%
\begin{eqnarray}
\label{Wxypar}  
W^{\nu\pi}\left(x_{\parallel},y_{\parallel}\right)&=&\frac{i}{2}\int\frac{d^{3}p_{\parallel}}{\left(2\pi\right)^{3}}\frac{e^{-ip_{\parallel}\cdot\left(x_{\parallel}-y_{\parallel}\right)}}{\Gamma}\Biggl\{\eta_{\parallel}^{\pi\nu}\Biggl[p_{\parallel}^{2}-\frac{\left[\left(k_{AF}\right)_{\parallel}\cdot p_{\parallel}\right]^{2}}{2p_{\parallel}^{2}}+\frac{3\left[\left(k_{AF}\right)^{3}\right]^{2}}{2}\Biggr]\nonumber\\
&
&+\frac{2\left[\left(k_{AF}\right)_{\parallel}\cdot p_{\parallel}\right]}{p_{\parallel}^{2}}\left[p_{\parallel}^{\pi}\left(k_{AF}\right)_{\parallel}^{\nu}+\left(k_{AF}\right)_{\parallel}^{\pi}p_{\parallel}^{\nu}\right]-2\left(k_{AF}\right)_{\parallel}^{\pi}\left(k_{AF}\right)_{\parallel}^{\nu}-i\left(k_{AF}\right)^{3}\epsilon^{\pi\nu\tau 3}p_{\parallel\tau}\Biggr\} \ ,
\end{eqnarray}
%%%%%%%%%%%%%%%%%%%%%
%%%%%%%%%%%%%%%%%%%%%%
\begin{eqnarray}
\label{Ixypara}    
I^{\nu}\left(x_{\parallel}\right)=\int d^{4}y \  f^{\nu}_{\ \mu}\left(y,x_{\parallel}\right)J^{\mu}\left(y\right) \ ,
\end{eqnarray}
%%%%%%%%%%%%%%%%%%%%%%%
with the definition
%%%%%%%%%%%%%%%%%%%%%%
\begin{eqnarray}
\label{fnmyxp}   
f^{\nu}_{\ \mu}\left(y,x_{\parallel}\right)&=&\frac{1}{2}\int\frac{d^{3}p_{\parallel}}{\left(2\pi\right)^{3}} \ e^{-ip_{\parallel}\cdot\left(x_{\parallel}-y_{\parallel}\right)}\Biggl\{\epsilon_{3 \ \ \mu}^{\ \nu\gamma} \ p_{\parallel\gamma}\Biggl[-1+\frac{1}{2}\Biggl(\frac{3\left[i\Gamma\mid y^{3}-a\mid - 1\right]}{p_{\parallel}^{2}}+\mid y^{3} -a\mid^{2}\Biggr)\nonumber\\
&
&\times\Biggl(\left[\left(k_{AF}\right)^{3}\right]^{2}+\frac{\left[\left(k_{AF}\right)_{\parallel}\cdot p_{\parallel}\right]^{2}}{p_{\parallel}^{2}}\Biggr)
+\frac{\left[i\Gamma\mid y^{3}-a\mid - 1\right]}{p_{\parallel}^{2}}\Biggl(i\left(k_{AF}\right)^{3}\left(y^{3}-a\right)\left[\left(k_{AF}\right)_{\parallel}\cdot p_{\parallel}\right]-2\left(k_{AF}^{2}\right)_{\parallel}\Biggr)\Biggr]\nonumber\\
&
&+\frac{\left[i\Gamma\mid y^{3}-a\mid - 1\right]}{p_{\parallel}^{2}}
\Biggl[2\left(k_{AF}\right)_{\mu}\epsilon_{3}^{\ \nu\gamma\alpha}p_{\parallel\gamma}+i\eta_{\mu 3}\left(\left(k_{AF}\right)_{\parallel}^{\nu}p_{\parallel}^{2}-p_{\parallel}^{\nu}\left[\left(k_{AF}\right)_{\parallel}\cdot p_{\parallel}\right]\right)\nonumber\\
&
&+i\left(k_{AF}\right)^{3}\left(\eta_{\parallel\mu}^{\nu}p_{\parallel}^{2}-p_{\parallel}^{\nu}p_{\parallel\mu}\right)\Biggr]
+\Biggl[\left(k_{AF}\right)_{\parallel}^{\nu}p_{\parallel\mu}-\eta_{\parallel\mu}^{\nu}\left[\left(k_{AF}\right)_{\parallel}\cdot p_{\parallel}\right]\Biggr]\left(y^{3}-a\right)\Biggr\}\frac{e^{i\Gamma\mid y^{3}-a\mid}}{\Gamma} \ ,
\end{eqnarray}
%%%%%%%%%%%%%%%%%%%
where $\left(k_{AF}\right)_{\parallel}^{\mu}=\left(\left(k_{AF}\right)^{0},\left(k_{AF}\right)^{1},\left(k_{AF}\right)^{2},0\right)$ stands for the background vector parallel to the plate.

Now, in the functional integral (\ref{fgplapar}) we perform the following translation,
%%%%%%%%%%%%%%%%%%%%
\begin{eqnarray}
 \label{translaB}   
B_{\nu}\left(x_{\parallel}\right)\rightarrow B_{\nu}\left(x_{\parallel}\right)+\int d^{3}y_{\parallel}V_{\nu\theta}\left(x_{\parallel},y_{\parallel}\right)I^{\theta}\left(y_{\parallel}\right) \ ,
\end{eqnarray}
%%%%%%%%%%%%%%%%%%%%%
where $V_{\nu\theta}\left(x_{\parallel},y_{\parallel}\right)$ is the function which inverts $W^{\nu\pi}\left(x_{\parallel},y_{\parallel}\right)$, in the sense that,
%%%%%%%%%%%%%%%%%%
\begin{eqnarray}
 \label{VFunction}   
\int d^{3}y_{\parallel}W^{\nu\pi}\left(x_{\parallel},y_{\parallel}\right)V_{\pi\theta}\left(y_{\parallel},z_{\parallel}\right)=\eta_{\parallel\theta}^{\nu} \ \delta^{3}\left(x_{\parallel}-z_{\parallel}\right) \ ,
\end{eqnarray}
%%%%%%%%%%%%%%%%%%%
namely,
%%%%%%%%%%%%%%%%%%%%%%
\begin{eqnarray}
V_{\pi\theta} \left(y_{\parallel},z_{\parallel}\right)&=&\int\frac{d^{3}p_{\parallel}}{\left(2\pi\right)^{3}}\Biggl\{-\frac{2i}{\Gamma}\Biggl[1+\frac{\left[\left(k_{AF}\right)_{\parallel}\cdot p_{\parallel}\right]^{2}}{2 p_{\parallel}^{4}}-\frac{\left[\left(k_{AF}\right)^{3}\right]^{2}}{2 p_{\parallel}^{2}}\Biggr]\eta_{\parallel\pi\theta}+\frac{2i}{\Gamma}\frac{\left[\left(k_{AF}\right)^{3}\right]^{2}}{p_{\parallel}^{4}}p_{\parallel\pi}p_{\parallel\theta}
\nonumber\\
&
&-\frac{4i}{\Gamma}\frac{\left(k_{AF}\right)_{\parallel\pi}\left(k_{AF}\right)_{\parallel\theta}}{p_{\parallel}^{2}}
+\frac{4i}{\Gamma}\frac{\left[\left(k_{AF}\right)_{\parallel}\cdot p_{\parallel}\right]}{p_{\parallel}^{4}}\left[p_{\parallel\pi}\left(k_{AF}\right)_{\parallel\theta}+\left(k_{AF}\right)_{\parallel\pi}p_{\parallel\theta}\right]\nonumber\\
&
&-\frac{2i}{\Gamma}\frac{i}{p_{\parallel}^{2}}\left(k_{AF}\right)^{3}\epsilon_{\pi\theta\rho 3}p_{\parallel}^{\rho}\Biggr\}e^{-ip_{\parallel}\cdot\left(y_{\parallel}-z_{\parallel}\right)} \ .
 \label{VFunction2}   
\end{eqnarray}
%%%%%%%%%%%%%%%%%%%%%%%
Resulting in,
%%%%%%%%%%%%%%%%%%%%
\begin{eqnarray}
\label{zbarfinal}    
{\bar{Z}}\left[J\right]={\bar{Z}}\left[0\right]\exp\left[-\frac{i}{2}\int d^{4}x \  d^{4}y \ J^{\mu}\left(x\right){\bar{D}}_{\mu\nu}\left(x,y\right)J^{\nu}\left(y\right)\right] \ ,
\end{eqnarray}
%%%%%%%%%%%%%%%%%%%%%%
where we identified the function, 
%%%%%%%%%%%%%%%%%%
\begin{eqnarray}
\label{propplate1}   
{\bar{D}}_{\mu\nu}\left(x,y\right)=-\int d^{3}\omega_{\parallel}d^{3}z_{\parallel}f^{\alpha}_{\ \mu}\left(x,\omega_{\parallel}\right)V_{\alpha\theta} \left(\omega_{\parallel},z_{\parallel}\right)
f^{\theta}_{\ \nu}\left(y,z_{\parallel}\right) \ .
\end{eqnarray}
%%%%%%%%%%%%%%%%%

Substituting (\ref{fnmyxp}) and (\ref{VFunction2}) in (\ref{propplate1}) and performing some manipulations, we obtain
%%%%%%%%%%%%%%%%%%%%
\begin{eqnarray}
 \label{prpoplate2}   
{\bar{D}}_{\mu\nu}\left(x,y\right)&=&-\frac{i}{2}\int\frac{d^{3}p_{\parallel}}{\left(2\pi\right)^{3}} \ e^{-ip_{\parallel}\cdot\left(x_{\parallel}-y_{\parallel}\right)}\Biggl[\left(\eta_{\parallel\mu\nu}-\frac{p_{\parallel\mu}p_{\parallel\nu}}{p_{\parallel}^{2}}\right)\left[1+M\left(x^{3},y^{3},p_{\parallel}\right)\right]+ \left[F_{1}\left(p_{\parallel}\right)\right]_{\mu\nu}\nonumber\\
&
&+\left[i\Gamma\mid y^{3}-a\mid-1\right]\left[F_{2}\left(p_{\parallel}\right)\right]_{\mu\nu}+\left(y^{3}-a\right)\left[F_{3}\left(p_{\parallel}\right)\right]_{\mu\nu}-\left[i\Gamma\mid x^{3}-a\mid-1\right]\left[F_{4}\left(p_{\parallel}\right)\right]_{\mu\nu}\nonumber\\
&
&-\left[i\Gamma\mid x^{3}-a\mid-1\right]\left[i\Gamma\mid y^{3}-a\mid-1\right]\left[F_{5}\left(p_{\parallel}\right)\right]_{\mu\nu}-\left(y^{3}-a\right)\left[i\Gamma\mid x^{3}-a\mid-1\right]\left[F_{6}\left(p_{\parallel}\right)\right]_{\mu\nu}\nonumber\\
&
&-\left(x^{3}-a\right)\left[F_{7}\left(p_{\parallel}\right)\right]_{\mu\nu}-\left(x^{3}-a\right)\left[i\Gamma\mid y^{3}-a\mid-1\right]\left[F_{8}\left(p_{\parallel}\right)\right]_{\mu\nu}\nonumber\\
&
&-\left(x^{3}-a\right)\left(y^{3}-a\right)\left[F_{9}\left(p_{\parallel}\right)\right]_{\mu\nu}\Biggr]\frac{
e^{i\Gamma\left(\mid x^{3}-a\mid +\mid y^{3}-a\mid\right)}}{\Gamma} \ ,
\end{eqnarray}
%%%%%%%%%%%%%%%%%%%%
where 
%%%%%%%%%%%%%%%%%%%
\begin{eqnarray}
 \label{funcM}   
M\left(x^{3},y^{3},p_{\parallel}\right)&=&-\frac{1}{2}\left[\left[\left(k_{AF}\right)^{3}\right]^{2}+\frac{\left[\left(k_{AF}\right)_{\parallel}\cdot p_{\parallel} \right]^{2}}{p_{\parallel}^{2}}\right]\Biggl[\frac{3}{p_{\parallel}^{2}}\Biggl(i\Gamma\Bigl(\mid x^{3}-a\mid +\mid y^{3}-a\mid\Bigr)-2\Biggr)\nonumber\\
&
&+\left(\mid x^{3}-a\mid^{2}+\mid y^{3}-a\mid^{2}\right)\Biggr]+\frac{i}{p_{\parallel}^{2}}\left(k_{AF}\right)^{3}\left[\left(k_{AF}\right)_{\parallel}\cdot p_{\parallel}\right]\Biggl[i\Gamma\Bigl(\mid x^{3}-a\mid\left(x^{3}-a\right)\nonumber\\
&
&+\mid y^{3}-a\mid\left(y^{3}-a\right)\Bigr)-\left(x^{3}-y^{3}\right)\Biggr]+\frac{2}{p_{\parallel}^{2}}\left(k_{AF}^{2}\right)_{\parallel}\left[i\Gamma\Bigl(\mid x^{3}-a\mid +\mid y^{3}-a\mid\Bigr)-2\Biggr)\right]\nonumber\\
&
&+\frac{\left[\left(k_{AF}\right)_{\parallel}\cdot p_{\parallel} \right]^{2}}{2p_{\parallel}^{4}}-\frac{\left[\left(k_{AF}\right)^{3}\right]^{2}}{2p_{\parallel}^{2}} \ ,
\end{eqnarray}
%%%%%%%%%%%%%%%%%%%

%%%%%%%%%%%%%%%%%%%%
\begin{eqnarray}
 \label{F1}   
\left[F_{1}\left(p_{\parallel}\right)\right]_{\mu\nu}&=&\frac{2}{p_{\parallel}^{4}}\Bigl\{\eta_{\parallel\mu\nu}\Bigl[p_{\parallel}^{2}\left(k_{AF}^{2}\right)_{\parallel}-\left[\left(k_{AF}\right)_{\parallel}\cdot p_{\parallel}\right]^{2}\Bigr]+\left[\left(k_{AF}\right)_{\parallel}\cdot p_{\parallel}\right]\Bigl[p_{\parallel\mu}\left(k_{AF}\right)_{\parallel\nu}+\left(k_{AF}\right)_{\parallel\mu}p_{\parallel\nu}\Bigr]\nonumber\\
&
&-p_{\parallel}^{2}\left(k_{AF}\right)_{\parallel\mu}\left(k_{AF}\right)_{\parallel\nu}-\left(k_{AF}^{2}\right)_{\parallel}p_{\parallel\mu}p_{\parallel\nu}\Bigr\}+\frac{i}{p_{\parallel}^{2}}\left(k_{AF}\right)^{3}\epsilon_{\mu\nu\lambda 3}p_{\parallel}^{\lambda} \ ,
\end{eqnarray}
%%%%%%%%%%%%%%%%%%%%

%%%%%%%%%%%%%%%%%%%%
\begin{eqnarray}
\label{F2}   
\left[F_{2}\left(p_{\parallel}\right)\right]_{\mu\nu}&=&-\frac{1}{p_{\parallel}^{4}}\Bigl\{2\left(k_{AF}\right)_{\nu}\Bigl[p_{\parallel}^{2}\left(k_{AF}\right)_{\parallel\mu}-\left[\left(k_{AF}\right)_{\parallel}\cdot p_{\parallel}\right]p_{\parallel\mu}\Bigr]-ip_{\parallel}^{2} \eta_{\nu 3}  \epsilon_{\mu\alpha\gamma 3}\left(k_{AF}\right)_{\parallel}^{\alpha}p_{\parallel}^{\gamma}-\left(k_{AF}\right)^{3}\Bigl[ip_{\parallel}^{2}\epsilon_{\mu\nu\gamma 3}p_{\parallel}^{\gamma}\nonumber\\
&
&+\eta_{\nu 3}\Bigl(p_{\parallel}^{2}\left(k_{AF}\right)_{\parallel\mu}-\left[\left(k_{AF}\right)_{\parallel}\cdot p_{\parallel}\right]p_{\parallel\mu}\Bigr)\Bigr]-\left[\left(k_{AF}\right)^{3}\right]^{2}\Bigl(\eta_{\parallel\mu\nu}p_{\parallel}^{2}-p_{\parallel\mu}p_{\parallel\nu}\Bigr)\Bigr\} \ ,
\end{eqnarray}
%%%%%%%%%%%%%%%%%%%%%

%%%%%%%%%%%%%%%%%%%%%%%
\begin{eqnarray}
\label{F3}   
\left[F_{3}\left(p_{\parallel}\right)\right]_{\mu\nu}&=& -\frac{1}{p_{\parallel}^{2}}\Bigl\{-p_{\parallel\nu}\epsilon_{\mu\alpha\gamma 3}\left(k_{AF}\right)_{\parallel}^{\alpha}p_{\parallel}^{\gamma}+\left[\left(k_{AF}\right)_{\parallel}\cdot p_{\parallel}\right]\epsilon_{\mu\nu\gamma 3}p_{\parallel}^{\gamma}+i\left(k_{AF}\right)^{3}\Bigl[\left(k_{AF}\right)_{\parallel\mu}p_{\parallel\nu}\nonumber\\
&
&-\left[\left(k_{AF}\right)_{\parallel}\cdot p_{\parallel}\right]\eta_{\parallel\mu\nu}\Bigr]\Bigr\} \ ,
\end{eqnarray}
%%%%%%%%%%%%%%%%%%%%%%

%%%%%%%%%%%%%%%%%%%%
\begin{eqnarray}
\label{F4}   
\left[F_{4}\left(p_{\parallel}\right)\right]_{\mu\nu}&=&-\frac{1}{p_{\parallel}^{4}}\Bigl\{-2\left(k_{AF}\right)_{\mu}\Bigl[p_{\parallel}^{2}\left(k_{AF}\right)_{\parallel\nu}-\left[\left(k_{AF}\right)_{\parallel}\cdot p_{\parallel}\right]p_{\parallel\nu}\Bigr]-ip_{\parallel}^{2} \eta_{\mu 3}  \epsilon_{\nu\alpha\lambda 3}\left(k_{AF}\right)_{\parallel}^{\alpha}p_{\parallel}^{\lambda}+\left(k_{AF}\right)^{3}\Bigl[ip_{\parallel}^{2}\epsilon_{\mu\nu\lambda 3}p_{\parallel}^{\lambda}\nonumber\\
&
&+\eta_{\mu 3}\Bigl(p_{\parallel}^{2}\left(k_{AF}\right)_{\parallel\nu}-\left[\left(k_{AF}\right)_{\parallel}\cdot p_{\parallel}\right]p_{\parallel\nu}\Bigr)\Bigr]+\left[\left(k_{AF}\right)^{3}\right]^{2}\Bigl(\eta_{\parallel\mu\nu}p_{\parallel}^{2}-p_{\parallel\mu}p_{\parallel\nu}\Bigr)\Bigr\} \ ,
\end{eqnarray}
%%%%%%%%%%%%%%%%%%%%%

%%%%%%%%%%%%%%%%%%%%%%%%%
\begin{eqnarray}
\label{F5}    
\left[F_{5}\left(p_{\parallel}\right)\right]_{\mu\nu}&=&-\frac{1}{p_{\parallel}^{4}}\Bigl\{\eta_{\mu 3}\eta_{\nu 3}\Bigl[p_{\parallel}^{2}\left(k_{AF}^{2}\right)_{\parallel}-\left[\left(k_{AF}\right)_{\parallel}\cdot p_{\parallel}\right]^{2}\Bigr]+\left(k_{AF}\right)^{3}\Bigl[\eta_{\mu 3}\Bigl(p_{\parallel}^{2}\left(k_{AF}\right)_{\parallel\nu}-\left[\left(k_{AF}\right)_{\parallel}\cdot p_{\parallel}\right]p_{\parallel\nu}\Bigr)\nonumber\\
&
&+\eta_{\nu 3}\Bigl(p_{\parallel}^{2}\left(k_{AF}\right)_{\parallel\mu}-\left[\left(k_{AF}\right)_{\parallel}\cdot p_{\parallel}\right]p_{\parallel\mu}\Bigr)\Bigr]+\left[\left(k_{AF}\right)^{3}\right]^{2}\Bigl(\eta_{\parallel\mu\nu}p_{\parallel}^{2}-p_{\parallel\mu}p_{\parallel\nu}\Bigr)\Bigr\} \ ,
\end{eqnarray}
%%%%%%%%%%%%%%%%%%%%%%%%

%%%%%%%%%%%%%%%%%%%%%%%%%%
\begin{eqnarray}
\label{F6}   
\left[F_{6}\left(p_{\parallel}\right)\right]_{\mu\nu}&=&\frac{i}{p_{\parallel}^{2}}\Bigl[\eta_{\mu 3}\Bigl(\left(k_{AF}^{2}\right)_{\parallel}p_{\parallel\nu}-\left[\left(k_{AF}\right)_{\parallel}\cdot p_{\parallel}\right]\left(k_{AF}\right)_{\parallel\nu}\Bigr)+\left(k_{AF}\right)^{3}\Bigl(\left(k_{AF}\right)_{\parallel\mu}p_{\parallel\nu}-\left[\left(k_{AF}\right)_{\parallel}\cdot p_{\parallel}\right]\eta_{\parallel\mu\nu}\Bigr)\Bigr] \ ,
\end{eqnarray}
%%%%%%%%%%%%%%%%%%%%%

%%%%%%%%%%%%%%%%%%%%%%%%%
\begin{eqnarray}
\label{F7}   
\left[F_{7}\left(p_{\parallel}\right)\right]_{\mu\nu}&=&-\frac{1}{p_{\parallel}^{2}}\Bigl[p_{\parallel\mu}\epsilon_{\nu\alpha\lambda 3}\left(k_{AF}\right)_{\parallel}^{\alpha}p_{\parallel}^{\lambda}+\left[\left(k_{AF}\right)_{\parallel}\cdot p_{\parallel}\right]\epsilon_{\mu\nu\lambda 3}p_{\parallel}^{\lambda}+i\left(k_{AF}\right)^{3}\Bigl(p_{\parallel\mu}\left(k_{AF}\right)_{\parallel\nu}\nonumber\\
&
&-\left[\left(k_{AF}\right)_{\parallel}\cdot p_{\parallel}\right]\eta_{\parallel\mu\nu}\Bigr)\Bigr] \ ,
\end{eqnarray}
%%%%%%%%%%%%%%%%%%%%%%%%%%

%%%%%%%%%%%%%%%%%%%%%%%%%%%
\begin{eqnarray}
 \label{F8}   
\left[F_{8}\left(p_{\parallel}\right)\right]_{\mu\nu}&=&-\frac{i}{p_{\parallel}^{2}} \Bigl[\eta_{\nu 3}\Bigl(\left(k_{AF}^{2}\right)_{\parallel}p_{\parallel\mu}-\left[\left(k_{AF}\right)_{\parallel}\cdot p_{\parallel}\right]\left(k_{AF}\right)_{\parallel\mu}\Bigr)+\left(k_{AF}\right)^{3}\Bigl(p_{\parallel\mu}\left(k_{AF}\right)_{\parallel\nu}-\left[\left(k_{AF}\right)_{\parallel}\cdot p_{\parallel}\right]\eta_{\parallel\mu\nu}\Bigr)\Bigr],
\end{eqnarray}
%%%%%%%%%%%%%%%%%%%%%%%%

%%%%%%%%%%%%%%%%%%%%%%%%%%
\begin{eqnarray}
 \label{F9}   
\left[F_{9}\left(p_{\parallel}\right)\right]_{\mu\nu}&=&\frac{1}{p_{\parallel}^{2}}\Bigl[-\eta_{\parallel\mu\nu}\left[\left(k_{AF}\right)_{\parallel}\cdot p_{\parallel}\right]^{2}-\left(k_{AF}^{2}\right)_{\parallel}p_{\parallel\mu}p_{\parallel\nu}+\left[\left(k_{AF}\right)_{\parallel}\cdot p_{\parallel}\right]\Bigl(p_{\parallel\mu}\left(k_{AF}\right)_{\parallel\nu}+\left(k_{AF}\right)_{\parallel\mu}p_{\parallel\nu}\Bigr)\Bigr] \ .
\end{eqnarray}
%%%%%%%%%%%%%%%%%%%%%%%%%%%

Substituting (\ref{zbarfinal}) and (\ref{fgen6}) in (\ref{fgen5}), the functional generator in the presence of a perfectly conducting plate becomes
%%%%%%%%%%%%%%%%%%%%%%%%%%%
\begin{eqnarray}
 \label{gfuntotal}   
{{Z}}_{C}\left[J\right]={{Z}}_{C}\left[0\right]\exp\left[-\frac{i}{2}\int d^{4}x \  d^{4}y \ J^{\mu}\left(x\right)\left({{D}}_{\mu\nu}\left(x,y\right)+{\bar{D}}_{\mu\nu}\left(x,y\right)\right)J^{\nu}\left(y\right)\right] \ .
\end{eqnarray}
%%%%%%%%%%%%%%%%%%%%%%%%%%
From expression (\ref{gfuntotal}), we can identify the propagator of the theory (\ref{lagEm}) in the presence of a conducting surface up to second order in the background vector, as follows
%%%%%%%%%%%%%%%%%%%%%
\begin{eqnarray}
 \label{prototal}   
{{D}}_{C}^{\mu\nu}\left(x,y\right)={{D}}^{\mu\nu}\left(x,y\right)+{\bar{D}}^{\mu\nu}\left(x,y\right) \ .
\end{eqnarray}
%%%%%%%%%%%%%%%%%%%%%%%

We can check the consistence of the results taking into account that the field solution produced by an external source is given by
%%%%%%%%%%%%
\begin{eqnarray}
\label{pfield}
A^{\beta}(x)=\int d^{4}y \ D_{C}^{\beta\rho}(x,y) J_{\rho}(y) \ .
\end{eqnarray}
%%%%%%%%%%%%%%%%
Substituting Eq. (\ref{pfield}) into (\ref{condition1}) we rewrite the conducting 
plate condition as follows
%%%%%%%%%%%%%
\begin{eqnarray}
\label{pfield11}
\int d^{4}y\left[\epsilon_{3\nu\alpha\beta}\frac{\partial D_{C}^{\beta\rho}(x,y)}
{\partial x_{\alpha}}\right]J_{\rho}(y)|_{x^{3}=a}=0\cr\cr
\Rightarrow \epsilon_{3\nu\alpha\beta}\frac{\partial D_{C}^{\beta\rho}(x,y)}
{\partial x_{\alpha}}\Big|_{x^{3}=a}=0\ .
\end{eqnarray}
 Substituting Eq. (\ref{prototal}) into (\ref{pfield11}) and  using Eqs. (\ref{prop2}), 
(\ref{intp31}) to (\ref{intp313}), and (\ref{prpoplate2}) to (\ref{F9}), after some manipulations, we can verify that the conducting plate condition (\ref{pfield11}) is satisfied.
%%%%%%%%%%%%%%%%

We notice that the propagator (\ref{prototal}) is composed by the sum of the free propagator propagator (\ref{prop2}) with the correction (\ref{prpoplate2}) which accounts for the presence of the conducting plate.  In the limit $\left(k_{AF}\right)^{\mu}\rightarrow 0$, the propagator (\ref{prpoplate2}) reduces to the same one as that found with the Maxwell electrodynamics in the presence of a conducting surface.

%%%%%%%%%
\section{CHARGE-PLATE INTERACTION}
\label{III}
%%%%%%%%%

Using the obtained propagator in the previous section, here we consider the interaction energy between a steady point-like charge and the conducting plate, which is given by \citep{FABFEB,OliveiraBorgesAFF,FABAAN1,BorgesBarone22,plane1,plane2,LHCBFABplate,LHCBFABplate2,LHCBAFFSM}
%%%%%%%%%%%%%%%%%%%%%%%
\begin{equation}
E=\frac{1}{2T}\int d^{4}x\ d^{4}y\ J^{\mu}\left(x\right){\bar{D}}_{\mu\nu}\left(x,y\right)J^{\nu}\left(y\right)\ ,\label{energy}
\end{equation}
%%%%%%%%%%%%%%%%%%%%%%%
where $T$ is a time interval, and it is implicit the limit $T\to\infty$
at the end of the calculations.

\begin{figure}[!h]
\centering 
\includegraphics[scale=0.40]{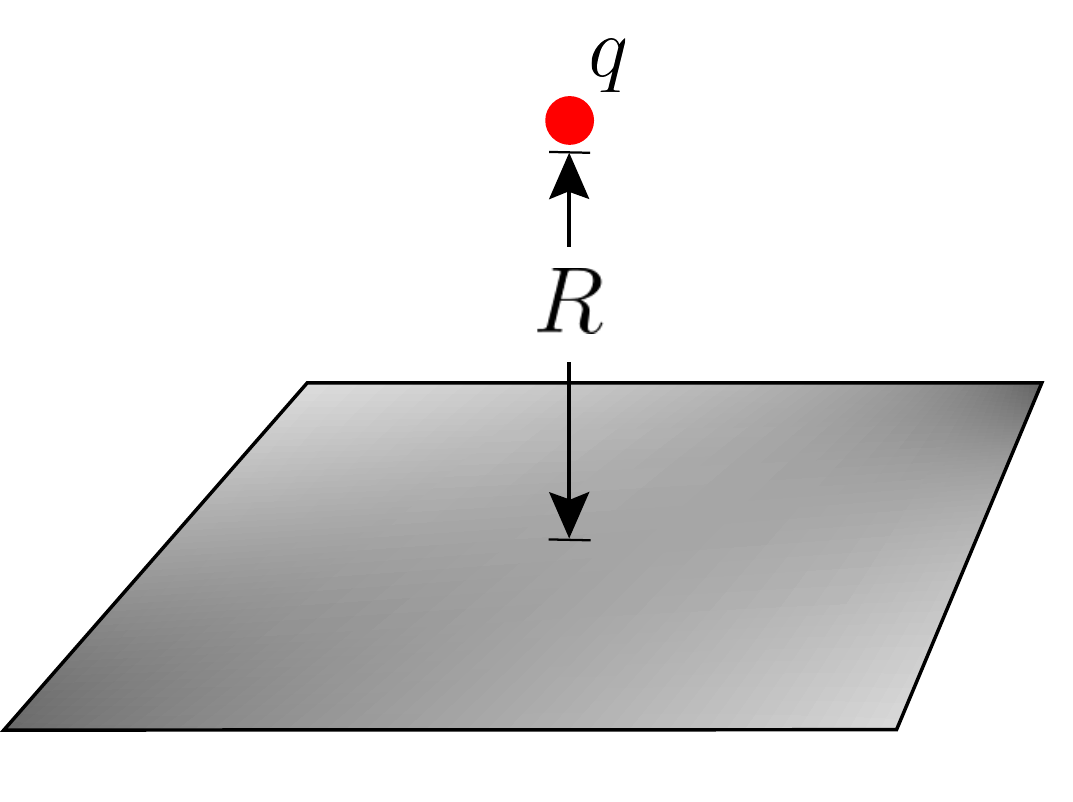}
\caption{Charge perpendicular to the plate.}
\label{figu2}
\end{figure}

With no loss of generality, we choose a point-like charge placed at the position ${\bf b}=\left(0,0,b\right)$, perpendicular to the plate, as shown in Fig. \ref{figu2}. The corresponding external source reads 
\begin{equation}
J^{\mu}\left(x\right)=q\eta^{\mu0}\delta^{3}\left({\bf x}-{\bf b}\right)\ ,\label{source1}
\end{equation}
where the parameter $q$ is the electric charge.
%%%%%%%%%%%%%%%%%%%%%%%%%
%%%%%%%%%%%%%%%%%%%%%%%%%%%%%%%%%

Substituting Eqs. (\ref{source1}) and (\ref{prpoplate2}) in (\ref{energy}), using the expressions (\ref{funcM}), (\ref{F1}), (\ref{F2}),(\ref{F3}), (\ref{F4}), (\ref{F5}), (\ref{F6}), (\ref{F7}), (\ref{F8}), (\ref{F9}), performing the integrals in the following order: $d^{3}{\bf x}$, $d^{3}{\bf y}$,
$dx^{0}$, $dp^{0}$, $dy^{0}$ and carrying out some manipulations, we obtain 
%%%%%%%%%%%%%%%%%%%%%%%%
\begin{eqnarray}
 \label{EPC1}
 E_{PC}&=&-\frac{q^{2}}{16\pi^{2}}\int d^{2}{\bf {p}}_{\parallel}\frac{e^{-2R\sqrt{{\bf {p}}_{\parallel}^{2}}}}{\sqrt{{\bf {p}}_{\parallel}^{2}}}\Biggl\{1-\frac{1}{{\bf{p}}_{\parallel}^{2}}\Biggl(\frac{1}{2}+R\sqrt{{\bf {p}}_{\parallel}^{2}}\Biggr)\Biggl[3\left[\left(k_{AF}\right)^{3}\right]^{2}\nonumber\\
 &
 &-\frac{3\left[\left({\bf{k}}_{AF}\right)_{\parallel}\cdot {\bf{p}}_{\parallel} \right]^{2}}{{\bf{p}}_{\parallel}^{2}}+4\left({\bf{k}}_{AF}^{2}\right)_{\parallel}\Biggr]-2R^{2}\left[\left(k_{AF}\right)^{3}\right]^{2}\Biggr\} \ ,
\end{eqnarray}
%%%%%%%%%%%%%%%%%%%%%%%%%%%%%
where $R=\mid b-a\mid$ is the distance between the plate and the charge. The subscript $PC$ indicates that we have the interaction energy between the plate and the charge. This result can be further simplified by using polar coordinates and integrating over the solid angle,
%%%%%%%%%%%%%%%%%%%%%%
\begin{eqnarray}
\label{EPC2}   
 E_{PC}&=&-\frac{q^{2}}{16\pi}\Biggl\{2\Bigl[1-2R^{2}\left[\left(k_{AF}\right)^{3}\right]^{2}\Bigr]\int_{0}^{\infty}dp \ e^{-2Rp}-\Bigl[6\left[\left(k_{AF}\right)^{3}\right]^{2}+5\left({\bf{k}}_{AF}^{2}\right)_{\parallel}\Bigr]\nonumber\\
 &
 &\times\Biggl(R\int_{0}^{\infty}dp \ \frac{e^{-2Rp}}{p}+\frac{1}{2}\int_{0}^{\infty}dp \ \frac{e^{-2Rp}}{p^{2}}\Biggr)\Biggr\} \ .
\end{eqnarray}
%%%%%%%%%%%%%%%%%%%%%%%
The first integral on the right-hand side of Eq. (\ref{EPC2}) is given by
%%%%%%%%%%%%%%%%%%%
\begin{equation}
\label{intEP1}  
\int_{0}^{\infty}dp \ e^{-2Rp}=\frac{1}{2R} \ .
\end{equation}
%%%%%%%%%%%%%%%%%%%%
However, the remaining integrals are divergent. They can be regularized by inserting a parameter $\epsilon$, as follows
%%%%%%%%%%%%%%%%%  
\begin{eqnarray}
\label{contribution12}
\int_{0}^{\infty} dp \frac{e^{-2Rp}}{p}&=&\lim_{\epsilon\rightarrow 0}\int_{\epsilon}^{\infty} dp \frac{e^{-2Rp}}{p}=\lim_{\epsilon\rightarrow 0}\left[Ei\left(1,2R\epsilon\right)\right] \ , \\
\int_{0}^{\infty} dp \frac{e^{-2Rp}}{p^{2}}&=&\lim_{\epsilon\rightarrow 0}\int_{\epsilon}^{\infty} dp \frac{e^{-2Rp}}{p^{2}}=\lim_{\epsilon\rightarrow 0}\left[\frac{e^{-2R\epsilon}}{\epsilon}-2R[Ei\left(1,2R\epsilon\right)\right] \ ,
\end{eqnarray}
%%%%%%%%%%%%%%%%%
where  $Ei\left(n, s\right)$ is the exponential integral function \cite{Arfken}, defined by 
%%%%%%%%%%%%%%%%
\begin{equation}
\label{Ei}
Ei\left(n, s\right)=\int^{\infty}_{1}\frac{e^{-ts}}{t^{n}}dt \ \ \ {\Re}\left(s\right)>0 \ , \ n = 0, 1, 2, \cdots \ .
\end{equation}
%%%%%%%%%%%%%%%%%%

Therefore, the interaction energy reads,
%%%%%%%%%%%%%%%%%%
\begin{eqnarray}
\label{EPC3}  
 E_{PC}&=&-\frac{q^{2}}{16\pi}\Bigg\{\frac{1}{R}\Bigl[1-2R^{2}\left[\left(k_{AF}\right)^{3}\right]^{2}\Bigr]-\Bigl[6\left[\left(k_{AF}\right)^{3}\right]^{2}+5\left({\bf{k}}_{AF}^{2}\right)_{\parallel}\Bigr]\lim_{\epsilon\rightarrow 0}\left(\frac{e^{-2R\epsilon}}{2\epsilon}\right)\Biggr\} \ .
\end{eqnarray}
%%%%%%%%%%%%%%%%%%%
Using the fact that
%%%%%%%%%%%%%%%%%%%
\begin{eqnarray}
 \label{aprox1}   
\frac{e^{-2R\epsilon}}{2\epsilon}\cong \frac{1}{2\epsilon}-R+{\cal{O}}\left(\epsilon\right) \ ,
\end{eqnarray}
%%%%%%%%%%%%%%%%%%
we arrive at
%%%%%%%%%%%%%%%%%%
\begin{eqnarray}
\label{EPC4}    
E_{PC}&=&-\frac{q^{2}}{16\pi}\Bigg\{\frac{1}{R}\Bigl[1-2R^{2}\left[\left(k_{AF}\right)^{3}\right]^{2}\Bigr]-\Bigl[6\left[\left(k_{AF}\right)^{3}\right]^{2}+5\left({\bf{k}}_{AF}^{2}\right)_{\parallel}\Bigr]\left(-R+\lim_{\epsilon\rightarrow 0}\frac{1}{2\epsilon}\right)\Biggr\} \ .
\end{eqnarray}
%%%%%%%%%%%%%%%%%%
Now, in Eq. (\ref{EPC4}), we can neglect the divergent term that does not depend on the distance $R$, as it does not contribute to the interaction force between the charge and the conducting plate. Therefore, we obtain
%%%%%%%%%%%%%%%%%%
\begin{eqnarray}
\label{EPC5}   
E_{PC}&=&-\frac{q^{2}}{16\pi}\Biggl[\frac{1}{R}+R\Bigl(5\left({\bf{k}}_{AF}^{2}\right)_{\parallel}+4\left[\left(k_{AF}\right)^{3}\right]^{2}\Bigr)\Biggr] \ .
\end{eqnarray}
%%%%%%%%%%%%%%%%
Equation (\ref{EPC5}) is a perturbative result up to second order in the background vector for the interaction energy of a point-like charge and a perfect mirror mediated by the model (\ref{lagEm}). The first term on the right-hand side reproduces the result obtained in standard Maxwell electrodynamics, while the remaining contributions are corrections due to Lorentz symmetry breaking.

From the Eq. (\ref{EPC5}) we obtain the interaction force between the conducting plate and the charge
%%%%%%%%%%%%%%%%%%%%
\begin{eqnarray}
 \label{FPC1}   
F_{PC}=-\frac{\partial E_{PC} }{\partial R}=\frac{q^{2}}{16\pi}\Biggl[-\frac{1}{R^{2}}+5\left({\bf{k}}_{AF}^{2}\right)_{\parallel}+4\left[\left(k_{AF}\right)^{3}\right]^{2}\Biggr] \ .
\end{eqnarray}
%%%%%%%%%%%%%%%%%%%%%%%
In the first contribution, we have the usual Coulomb interaction between the charge $q$ and its image, placed at a distance $2R$ apart. The second and third terms are corrections imposed by the Lorentz-violating parameter up to second order.

In Eq. (10) of Ref. \cite{CL1} we have the interaction energy between two stationary point-like charges for the model (\ref{lagEm}) up to second order in $\left(k_{AF}\right)^{\mu}$. From this expression, we can obtain the interaction force for the specific case where we have two opposite charges, $q_{1}=q$ and $q_{2}=-q$ placed at a distance $2R$ apart. In this setup, we obtain that
%%%%%%%%%%%%%%%%%%%%
\begin{eqnarray}
\label{FCC1}   
F_{CC}=\frac{q^{2}}{16\pi}\Biggl[-\frac{1}{R^{2}}+6\left({\bf{k}}_{AF}^{2}\right)_{\parallel}+4\left[\left(k_{AF}\right)^{3}\right]^{2}\Biggr] \ ,
\end{eqnarray}
%%%%%%%%%%%%%%%%%%%%%%%
where the sub-index $CC$ means that we have the interaction between two point-like charges.

First, considering a background vector with no null spatial components parallel to the mirror, we observe that Eq. (\ref{FPC1}) is different from Eq. (\ref{FCC1}). Therefore, the image method is not valid for the CPT-odd gauge sector of the minimal SME up to second order in $\left(k_{AF}\right)^{\mu}$ for the conducting plate condition (\ref{condition1}). In contrast, the image method remains valid for the CPT-even gauge sector of the minimal SME \cite{LHCBFABplate}. However, for the nonminimal Lorentz-violating model considered in Ref. \cite{LHCBFABplate2}, the image method is not valid.

Now, considering a setup where the background vector just has the component perpendicular to the mirror, namely ${\bf{k}}_{AF}=\left(0, 0, \left(k_{AF}\right)^{3}\right)$, the expressions (\ref{FPC1}) and (\ref{FCC1}) become equivalent to each other. Thus, the image method is valid for the model (\ref{lagEm}) for the conducting plate condition (\ref{condition1}).
 
\begin{figure}[!h]
\centering 
\includegraphics[scale=0.40]{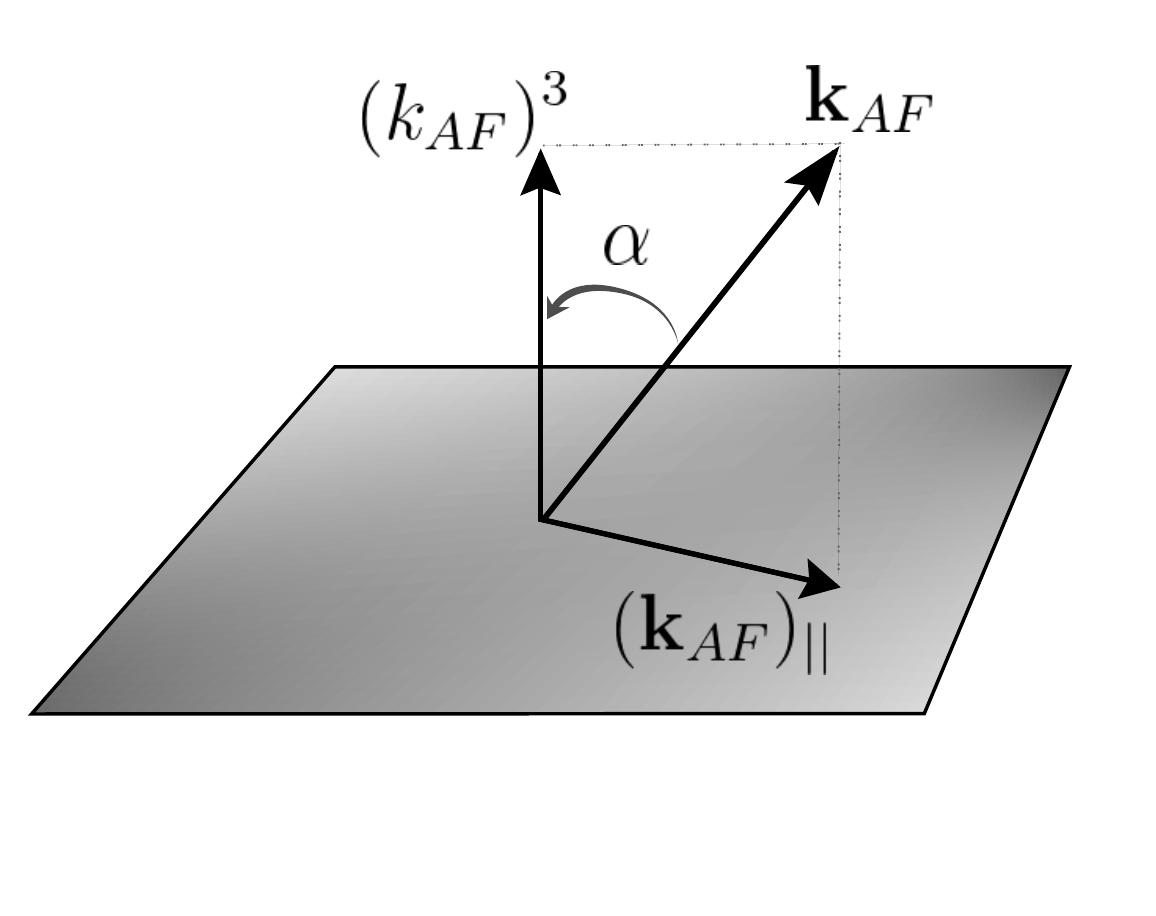}
\caption{The angle $\alpha$ between the normal to the mirror and the background vector.}
\label{figu3}
\end{figure}

 When we fix the distance between the charge and the mirror, from Eq. (\ref{EPC5}), we can obtain a torque on this physical system. In order to calculate this torque, we define as $0\leq \alpha\leq \pi$ the angle between the normal to the mirror and the background vector ${\bf{k}}_{AF}$, in such a way that, $\left({\bf{k}}_{AF}^{2}\right)_{\parallel}={\bf{k}}_{AF}^{2}\sin^{2}\left(\alpha\right) \ , \ \left[\left(k_{AF}\right)^{3}\right]^{2}={\bf{k}}_{AF}^{2}\cos^{2}\left(\alpha\right)$, as represented in Fig. \ref{figu3}. Thus, the expression (\ref{EPC5}) becomes 
%%%%%%%%%%%%%%%%%%%%%%%%%
%%%%%%%%%%%%%%%%%%%%%%%%%%%%%%%%%
%%%%%%%%%%%%%%%%%%%%%%
\begin{eqnarray}
\label{EPC6}    
E_{PC}\left(\alpha\right)&=&-\frac{q^{2}}{16\pi}\Biggl[\frac{1}{R}+ \left({\bf{k}}_{AF}^{2}\right)R\Bigl(4+\sin^{2}\left(\alpha\right)\Bigr)\Biggr] \ ,
\end{eqnarray}
%%%%%%%%%%%%%%%%%%%%%%%
and the torque can be obtained as follows
%%%%%%%%%%%%%%%%%%%%
\begin{eqnarray}
 \label{tor}   
\tau_{PC}=-\frac{\partial E_{PC}\left(\alpha\right)}{\partial\alpha}=\frac{q^{2}}{16\pi} \left({\bf{k}}_{AF}^{2}\right)R\sin\left(2\alpha\right) \ .
\end{eqnarray}
%%%%%%%%%%%%%%%%%%%%%%%

\begin{figure}[!h]
\centering 
\includegraphics[scale=0.45]{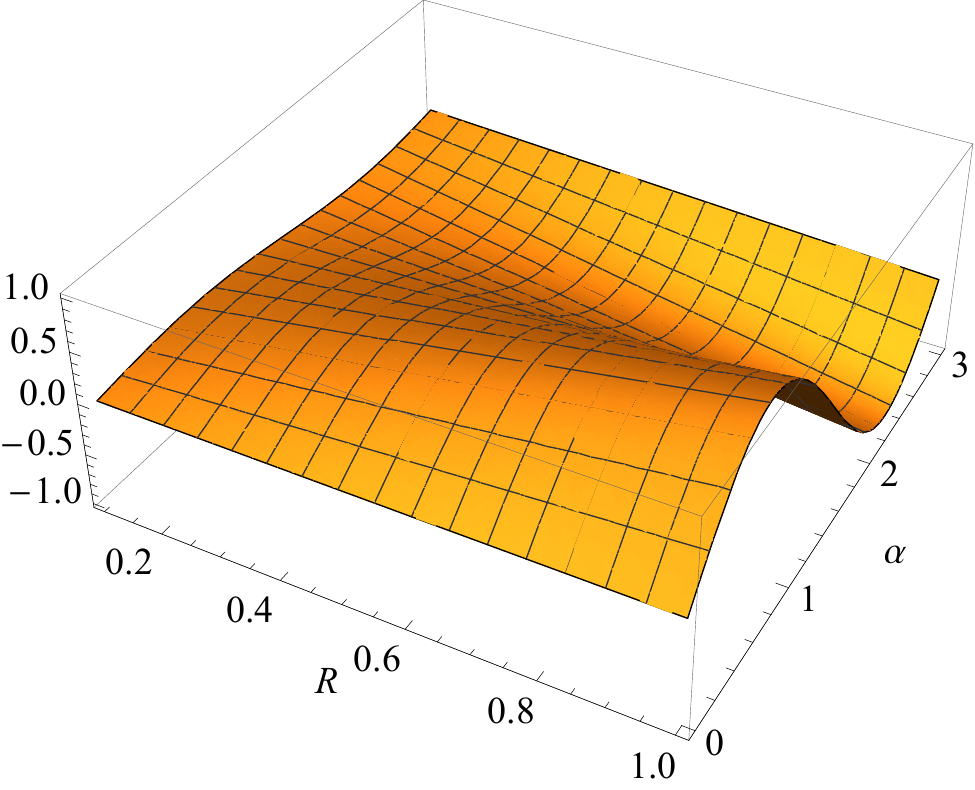}
\caption{The torque given in Eq. (\ref{tor}), multiplied by $\frac{16\pi}{q^{2}\left({\bf{k}}_{AF}^{2}\right)}$.}
\label{figu4}
\end{figure}

The torque (\ref{tor}) does not occur in the usual Maxwell electrodynamics, making it an exclusive Lorentz-violating effect up to second order in the background vector. If $\mid{\bf{k}}_{AF}\mid=0$ or $\alpha=0, \pi/2, \pi$ this effect is absent, and for $\alpha = \pi/4$ the torque reaches the maximum intensity. A similar effect was obtained in Ref. \cite{LHCBFABplate} for the CPT-even gauge sector of the minimal SME, as well as in Ref. \cite{LHCBFABplate2} for the nonminimal SME scenario.  In Fig. \ref{figu4}, the general behavior of the torque (\ref{tor}) is shown as a function of $R$ and $\alpha$.
%%%%%%%%%%%%%%%%%%%%%%%%%
%%%%%%%%%%%%%%%%%%%%%%%%%%%%%%%%%

The Lorentz violation parameter is strongly constrained from astrophysical test data to be of the order of $\mid{\bf{k}}_{AF}\mid\sim 10^{-42}$ GeV \cite{CFJ1}. Let's use this upper bound to obtain an estimate of the order of magnitude for the torque (\ref{tor}). Using the typical distance of atomic experiments in the vicinity of conductors, which is on the order of $R\sim 10^{-6}$ m, and the fundamental electronic charge $q\sim 1.60217\times 10^{-19}$ C, we obtain $\tau_{PC}\sim 10^{-89}$ Nm. However, the magnitude of the obtained effect is currently beyond the reach of existing technology for measurement.

%%%%%%%%%
\section{CONCLUSIONS AND FINAL REMARKS}
\label{IV}
%%%%%%%%%

In this paper, we have investigated some aspects of the CPT-odd gauge sector of the SME in the vicinity of a perfectly conducting plate (perfect mirror) and obtained perturbative results up to second order in the background vector. We computed the modified propagator for the gauge field due to the presence of the mirror and calculated the interaction energy, as well as the interaction force, between the mirror and a steady point-like charge. 

We showed that when the charge is placed in the vicinity of the mirror, a spontaneous torque arises in this physical system due to the orientation of the mirror with respect to the background vector. We concluded that, at present, this torque is beyond the reach of current measurements. However, this effect can be relevant in the theoretical context since it has no counterpart in usual Maxwell electrodynamics. Even in the experimental scope, maybe in the future, such an effect could be of some relevance in the context of QED in the presence of material media. Furthermore, the obtained results in this paper are new features regarding the CFJ electrodynamics.

We also showed that the image method for the CFJ electrodynamics is valid only in the setup where the background vector only has the component perpendicular to the mirror.

%%%%%%%%%%%%
\appendix
\section{Integrals (\ref{intp31})-(\ref{intp313})}
In this appendix, we provide some details on the calculation of the integrals
(\ref{intp31}), (\ref{intp312}), and (\ref{intp313}). For this task, we employ the well-known $i\varepsilon$-prescription to shift the poles away from the real axis in the complex plane of $p^{3}$. We can write
\[
\int\frac{dp^{3}}{\left(  2\pi\right)  }\frac{e^{ip^{3}(x^{3}-y^{3})}}{p^{2}%
}=-\int\frac{dp^{3}}{\left(  2\pi\right)  }\frac{e^{ip^{3}\left(  x^{3}%
		-y^{3}\right)  }}{\left(  p^{3}\right)  ^{2}-\Gamma^{2}-i\varepsilon}\ ,
\]
with the limit $\varepsilon\rightarrow0$ implicitly assumed. For $x^{3}%
-y^{3}>0$, we close the integration contour (infinite semicircle) in the upper half-plane, enclosing the simple pole at $\Gamma+i\varepsilon$. For $x^{3}-y^{3}<0$, we choose the integration contour in the lower half-plane, enclosing the simple pole at $-\Gamma-i\varepsilon$. Calculating the residue of the integrand at these poles and adding these contributions using the residue theorem, we obtain the result (\ref{intp31}).

For the integrals (\ref{intp312}) and (\ref{intp313}), the same procedure is followed. The only distinction is that these integrals have double and triple poles, respectively, which impacts the calculation of the residues.

%%%%%%%%%
\begin{acknowledgments}
D. M. Soares. thanks to Coordenação de Aperfeiçoamento de Pessoal de Nível Superior – Brasil (CAPES) for financial support.
\end{acknowledgments}
%%%%%%%%%

\subsubsection*{\bf{Data Availability Statement}} 

No Data associated in the manuscript.

\subsubsection*{\bf{Declarations}} 

\textbf{Conflict of interest} The authors declare no conflict of interest.

%%%%%%%%%

%%%%%%%%%


\begin{thebibliography}{99}


\bibitem{SME1} D. Colladay, V.A. Kostelecký, Phys. Rev. D \textbf{{55}}, 6760 (1997).


\bibitem{SME2}  S. R. Coleman and S. L. Glashow, Phys. Rev. D \textbf{{59}}, 116008 (1999).



\bibitem{SME3}  D. Colladay, V.A. Kostelecký, Phys. Rev. D \textbf{{58}}, 116002 (1998).


\bibitem{CFJ1} S.M. Carroll, G.B. Field and R. Jackiw, Phys Rev. D\textbf{{41}}, 1231 (1990).


\bibitem{CL1} L.H.C. Borges and A.F. Ferrari, Mod.Phys.Lett. A \textbf{{37}},  2250021 (2022).


\bibitem{CL2} Filipe S. Ribeiro, Pedro D. S. Silva, and Manoel M. Ferreira Jr., Phys. Rev. D \textbf{{107}}, 096018 (2023).


\bibitem{CL3} Eduardo Barredo-Alamilla, Luis F. Urrutia, and Manoel M. Ferreira Jr., Phys. Rev. D \textbf{{107}}, 096024 (2023)


\bibitem{CL4} Pedro D. S. Silva, Letícia Lisboa-Santos, Manoel M. Ferreira Jr., and Marco Schreck, Phys. Rev. D \textbf{{104}}, 116023 (2021).


\bibitem{CL5} M. M. Ferreira Jr., J.A. Helayël-Neto, C.M. Reyes, M. Schreck, and P.D.S. Silva, Phys. Lett. B \textbf{{804}}, 135379 (2020).


\bibitem{R1} J. Alfaro, A.A. Andrianov, M. Cambiaso, P. Giacconi, and R. Soldati, Int. J. Mod. Phys. A \textbf{{25}}, 3271 (2010).


\bibitem{R2} A. A. Andrianov, D. Espriu, P. Giacconi, and R. Soldati, J. High Energy Phys. \textbf{{09}} (2009).



\bibitem{R3} L. C. T. Brito, J. C. C. Felipe, A. Yu. Petrov, and A. P. Baêta Scarpelli, Int. J. Mod. Phys. A \textbf{{36}}, 2150033 (2021).


\bibitem{R4} T.R.S. Santos and R.F. Sobreiro,  Eur. Phys. J. C \textbf{{77}}, 903 (2017).



\bibitem{R5} J. C. C. Felipe, A. R. Vieira, A. L. Cherchiglia, A. P. Baêta Scarpelli, and M. Sampaio, , Phys. Rev. D \textbf{{89}}, 105034 (2014).


\bibitem{QED1} Y. M. P. Gomes and P. C. Malta,  Phys. Rev. D \textbf{{94}}, 025031 (2016).



\bibitem{QED2} C. Adam and F.R. Klinkhamer, Nucl. Phys. B \textbf{{607}}, 247 (2001); Nucl. Phys. B \textbf{{657}}, 214 (2003).



\bibitem{TP1}  R. Casana, M. M. Ferreira Jr., E. da Hora, and A. B. F. Neves, Eur. Phys. J. C \textbf{{74}}, 3064 (2014).



\bibitem{TP2}  R. Casana and L. Sourrouille, , Phys. Lett. B \textbf{{726}}, 488 (2013).



\bibitem{Ce0} V. A. Kostelecký and M. Mewes, Phys. Rev. D \textbf{{66}}, 056005 (2002).



\bibitem{Ce1} L.H.C. Borges, F.A. Barone, J.A. Helayël-Neto, Eur. Phys. J. C \textbf{{74}}, 2937 (2014).



\bibitem{Ce2} L.H.C. Borges, F.A. Barone, Braz. J. Phys. \textbf{{49}}, 571-582 (2019).



\bibitem{Ce3} L.H.C. Borges and F.A. Barone, Eur. Phys. J. C \textbf{{76}}, 64 (2016).


\bibitem{Ce4} A. Martín-Ruiz and C. A. Escobar, Phys. Rev. D \textbf{{94}}, 076010 (2016).



\bibitem{Ce5} M. Frank, I. Turan, Phys. Rev. D \textbf{{74}}, 033016 (2006).



\bibitem{Ce6} Rodolfo Casana, Manoel M. Ferreira Jr., Adalto R. Gomes, and Frederico E. P. dos Santos
Phys. Rev. D \textbf{{82}}, 125006 (2010).



\bibitem{Ce7}  Rodolfo Casana, Manoel M. Ferreira Jr., and Madson R. O. Silva Phys. Rev. D \textbf{{81}}, 105015
(2010).



\bibitem{Ce8} Rodolfo Casana, Manoel M. Ferreira Jr., Adalto R. Gomes, and Paulo R. D. Pinheiro Phys.
Rev. D \textbf{{80}}, 125040 (2009).



\bibitem{Ce9} V. A. Kostelecký and M. Mewes, Phys. Rev. Lett.\textbf{{87}}, 251304 (2001).




\bibitem{Ce10} V. A. Kostelecký and M. Mewes, Phys. Rev. Lett. \textbf{{66}}, 056005 (2002).




\bibitem{Ce11}  V. A. Kostelecký and M. Mewes, Phys.Rev. Lett. \textbf{{97}}, 140401 (2006).



\bibitem{Ce12} Quentin G. Bailey and V. Alan Kostelecký, Phys. Rev. D \textbf{{70}}, 076006 (2004).



\bibitem{Ce13} Frederico E.P. dos Santos and Manoel M. Ferreira, Symmetry \textbf{{10}}, 302 (2018).


\bibitem{FABFEB} F. A. Barone and F.E. Barone, Phys. Rev. D \textbf{{89}}, 065020 (2014).


\bibitem{GTFABFEB} G. T. Camilo, F. A. Barone and F. E. Barone, Phys. Rev. D \textbf{{87}}, 025011 (2013).


\bibitem{OliveiraBorgesAFF} H.L. Oliveira, L.H.C. Borges, F.E. Barone and F.A. Barone, Eur. Phys. J. C \textbf{{81}}, 558 (2021).


\bibitem{Milton} K.A. Milton, \textit{{The Casimir Effect, Physical
Manifestations of Zero-Point Energy, World Scientific, Singapore}}
(2001).

\bibitem{BorUM} M. Bordag, U. Mohideen, and V. M. Mostepanenko, Phys.
Rep. \textbf{{353}}, 1 (2001).

\bibitem{KimballA} Kimball A. Milton, J. Phys. A: Math. Gen.\textbf{{37}},
63916406 (2004).

\bibitem{BordKD} M. Bordag, K. Kirsten and D. Vassilevich, Phys.
Rev. D \textbf{{59}}, 085011 (1999).

\bibitem{NRVMH} N. Graham, R.L. Jaffe, V. Khemani, M. Quandt, M.
Scandurra and H. Weigel, Nucl. Phys. B \textbf{{645}}, 49 (2002).

\bibitem{NRVMMH2} N. Graham, R.L. Jaffe, V. Khemani, M. Quandt, M.
Scandurra and H. Weigel, Phys. Lett. B \textbf{{572}}, 196 (2003).

\bibitem{PsRj} P. Sundberg and R.L. Jaffe, Annals Phys. \textbf{{309}}, 442 (2004).

\bibitem{Caval} R.M. Cavalcanti, {[}arXiv:hep-th/0201150{]}.

\bibitem{FABFEB2} F. A. Barone and F. E. Barone, Eur. Phys. J. C \textbf{{74}}, 3113 (2014).


\bibitem{FABAAN1} F. A. Barone and A. A. Nogueira, Eur. Phys. J. C \textbf{{75}}, 339 (2015).

\bibitem{LW1} F.A. Barone and A.A. Nogueira, Int. J. Mod. Phys.:
Conf. Ser. \textbf{{41}}, 1660134, (2016).

\bibitem{LW2} M. Blazhyevska, J. of Phys. Stud. \textbf{{16}},
3001 (2012).

\bibitem{LW3} L.H.C. Borges, A.A. Nogueira, E.H. Rodrigues, F.A. Barone, Eur. Phys. J. C \textbf{{80}}, 1082 (2020).

\bibitem{BorgesBarone22} L.H.C. Borges and F.A. Barone, Phys. Lett. B 824, 136759 (2022).


\bibitem{plane1}  L. H. C. Borges, F. E. Barone, C. C. H. Ribeiro, H. L. Oliveira, R. L. Fernandes, F. A. Barone, Eur. Phys. J. C \textbf{{80}}, 238 (2020).


\bibitem{plane2} L.H.C. Borges, F.A. Barone, and H.L. Oliveira, Phys. Rev. D \textbf{{105}}, 025008 (2022).


\bibitem{CSE1} M.B. Cruz, E.R. Bezerra de Mello and A. Yu. Petrov,
Phys. Rev. D \textbf{{96}}, 045019 (2017).

\bibitem{CSE2} M.B. Cruz, E.R. Bezerra de Mello and A. Yu. Petrov,
Mod. Phys. Lett. A \textbf{{33}}, 1850115 (2018).

\bibitem{CSE3} M. Frank, I. Turan, Phys. Rev. D \textbf{{74}},
033016 (2006).

\bibitem{CSE4} A.F. Santos, F.C. Khanna, Phys. Lett. B \textbf{{762}}, 283 (2016).

\bibitem{CSE5} L. M. Silva, H. Belich, J. A. Helayel-Neto, arXiv:1605.02388 (2016).

\bibitem{CSE6} A. Martín-Ruiz, C.A. Escobar, Phys. Rev. D \textbf{{94}}, 076010 (2016).

\bibitem{CSE7} Dêivid R. da Silva and E. R. Bezerra de Mello, arXiv:200612924 (2020).

\bibitem{CSE8} A. Martín-Ruiz, C.A. Escobar, A.M. Escobar-Ruiz,
and O.J. Franca, Phys. Rev. D \textbf{{102}}, 015027 (2020).

\bibitem{CSE9}Amirhosein Mojavezi, Reza Moazzemi, Mohammad Ebrahim Zomorrodian, Nucl. Phys. B \textbf{{941}}, 145-157 (2019).

\bibitem{CSE10} M.B. Cruz, E.R. Bezerra de Mello, and A. Yu. Petrov,
Phys. Rev. D \textbf{{99}}, 085012 (2019).

\bibitem{CSE11}C. A. Escobar, Leonardo Medel, and A. Martín-Ruiz,
Phys. Rev. D \textbf{{101}}, 095011 (2020).

\bibitem{CSE12} Andrea Erdas, arXiv:200507830 (2020).

\bibitem{CSE13} M.A. Valuyan, Mod. Phys. Lett. A \textbf{{35}},
2050149 (2020).

\bibitem{CSE14} M.B. Cruz, E.R. Bezerra de Mello, and H. F. Santana Mota, Phys. Rev. D \textbf{{102}}, 045006 (2020).

\bibitem{CSE15} Massimo Blasone, Gaetano Lambiase, Luciano Petruzziello, and Antonio Stabile, Eur. Phys. J. C \textbf{{78}}, 976 (2018).

\bibitem{CSE16} Robson A. Dantas, Herondy F. Santana Mota  and Eugênio R. Bezerra de Mello, Universe \textbf{{9}}, 241 (2023). 

\bibitem{CSE17} Ar Rohim,  Apriadi Salim Adam and Arista Romadani, arXiv:2307.04448v1 [hep-th] (2023).


\bibitem{CSE18} M. C. Araújo, J. Furtado, R. V. Maluf, arXiv:2302.08836v1 [hep-th] (2023).


\bibitem{CSE19} E. R. Bezerra de Mello and M. B. Cruz, International Journal of Modern Physics A \textbf{{38}}, 2350062 (2023).


\bibitem{CSE20} A. Martín-Ruiz and C. A. Escobar, Phys. Rev. D \textbf{{95}}, 036011 (2017).


\bibitem{LHCBFABplate} L.H.C. Borges and F.A. Barone, Eur. Phys. J. C \textbf{{77}}, 693 (2017).

\bibitem{LHCBFABplate2} L.H.C. Borges and F.A. Barone, Braz. J. Phys. \textbf{{50}}, 647-657 (2020).


\bibitem{H1} V. A. Kostelecký and M. Mewes, Phys. Rev. D  \textbf{{88}}, 096006 (2013).


\bibitem{H2} V. A. Kostelecký and M. Mewes, Phys. Rev. D  \textbf{{80}}, 015020 (2009).


\bibitem{H3} V. Alan Kostelecký, Z. Li, Phys. Rev. D  \textbf{{99}}, 056016 (2019).

\bibitem{LHCBAFFFAB} L. H. C. Borges, A. F. Ferrari and F. A. Barone, Nucl. Phys. B \textbf{{954}}, 114974 (2020).

\bibitem{LHCBAFFSM} L. H. C. Borges and A. F. Ferrari, Nucl. Phys. B \textbf{{980}}, 115829 (2022).


\bibitem{MBor85} M. Bordag, D. Robaschik, E. Wieczorek, Quantum field theoretic
treatment of the casimir effect. Ann. Phys.  \textbf{{165}}, 192–213 (1985).

\bibitem{Arfken} G.B. Arfken and H. J. Weber, {\it{Mathematical Methods for Physicists}}, Academic Press (1995).


\end{thebibliography}
\end{document}